\pdfoutput=1
\documentclass[prd,
               twocolumn,
               amssymb,
               amsmath,
               eqsecnum,
               showpacs,
               letterpaper,
               superscriptaddress,
               altaffilletter,
               floatfix]
               {revtex4-1}

\usepackage{color}
\usepackage{graphicx}
\usepackage{latexsym}
\usepackage{float}
\usepackage{amsmath}
\usepackage{amssymb}
\usepackage{multirow}
\RequirePackage{fix-cm}

\begin{document}

\title{Reconstruction of Chirp Mass in the Search of Compact Binaries}

\author{V.~Tiwari}  
\affiliation{University of Florida, P.O.Box 118440, Gainesville, Florida, 32611, USA}
\affiliation{Cardiff University, Cardiff CF24 3AA, United Kingdom}
\author{S.~Klimenko}  
\affiliation{University of Florida, P.O.Box 118440, Gainesville, Florida, 32611, USA}
\author{V.~Necula}
\affiliation{University of Florida, P.O.Box 118440, Gainesville, Florida, 32611, USA}
\author{G.~Mitselmakher}
\affiliation{University of Florida, P.O.Box 118440, Gainesville, Florida, 32611, USA}

\begin{abstract}
Excess energy method is used in searches of gravitational waves (GWs) produced from sources with poorly modeled characteristics. It identifies GW events by searching for a coincidence appearance of excess 
energy in a GW detector network. While it is sensitive to a wide range of signal morphologies, 
the energy outliers can be populated by background noise events (background), 
thereby reducing the statistical confidence of a true signal. However, if the physics of the source is partially understood, weak model dependent constraints can be imposed to suppress the 
background. This letter presents a novel idea of using the reconstructed chirp mass along with two goodness of fit parameters for suppressing background when search is focused on GW produced from the compact 
binary coalescence.
\end{abstract}

\date[\relax]{Dated: \today }

\maketitle

%PACS numbers: 04.30.Db, 04.80.Nn, 07.05.Kf, 07.05.Rm, 95.30.Sf, 95.85.Sz
%(Some figures in this article are in colour only in the electronic version)

\section{Introduction}

Laser Interferometer Gravitational-Wave Observatory (LIGO) is a large-scale physics experiment targeting the first direct detection and study of gravitational waves from astrophysical sources~\cite{LIGO}. Two 
LIGO detectors in Livingston, LA and Hanford, WA have been upgraded to increase their sensitivity by an order of magnitude and started to take data in September 2015.

LIGO data is searched for GW signals by using matched filtering technique when GW waveforms of the source can be modeled. 
Furthermore, if source model is not understood or it is computationally or scientifically challenging to model waveforms, excess energy methods (aka `burst method') are used in the analysis. Coherent WaveBurst (cWB)
method~\cite{kymm2008} has been used to conduct multiple searches in 
the past~\cite{burstS52009, burstS62012, lvc2012}. This method identify GW events by searching for coincidence appearance of 
excess energy in a network of GW detectors. The cWB parameter space is not constrained, because of its ``\emph{eyes wide open approach}'' and allows for the detection of multiple types 
of signals. Because of that, the cWB method is affected by background noises, which 
can be suppressed by providing the search an additional 
discriminating power~\cite{PRD05}. This can be done by the application of weak model based constraints.

In the presented letter we describe the event selection criteria based on the reconstructed chirp mass (addressed as chirp cut from henceforth) in a burst search for compact binary coalescence (CBC) sources. 
Reconstructed chirp mass along with two goodness of fit parameters is shown to greatly suppresses the background while keeping the robustness of the search intact. This letter is organized as follows: 
In Section~\ref{algo} we describe the algorithm used for the reconstruction of the chirp mass; in Section~\ref{app} we discuss the application of the algorithm; results are discussed in Section~\ref{res} 
and paper is concluded in Section~\ref{conclusion}.

\section{Algorithm \label{algo}}

The frequency evolution of GW from a coalescing binary is given by Equation \ref{chirpmassfreq}
\begin{equation} \dot{f} = \frac{96}{5} \pi^{8/3}\left(\frac{\text{G}M_c}{\text{c}^3}\right)^{5/3}f^{11/3}, \label{chirpmassfreq}\end{equation}
where $M_c=(m_1m_2)^{3/5}/(m_1+m_2)^{1/5}$ is the chirp mass of the binary with component masses $m_1$ and $m_2$, G is the gravitational constant, and c is the speed of light \cite{ca2007}. Integrating 
Equation \ref{chirpmassfreq} with respect to time gives,
\begin{equation} \frac{96}{5}\pi^{8/3}\left(\frac{\text{G}M_\text{c}}{c^3}\right)^{5/3}t+\frac{3}{8}f^{-8/3}+C=0,\label{inteqchirpmassfreq}\end{equation}
where $C$ is the constant of integration. Furthermore, on identifying $x \equiv t$ and $y\equiv \frac{3}{8}f^{-8/3}$, the chirp mass can be calculated from the slope of the line fitted through the data points. 

In our case, line fitting is performed on a collection of time-frequency (TF) pixels extracted from an excess energy event appearing in the detector network data. The center of the pixels are promoted as the 
data points by including dimension of the pixels ($\Delta f$ and $\Delta t$) as errors in the $\chi^2$, defined as: 
\begin{equation} \chi^2 \equiv \sum_{i}\frac{\left( y_i-b(x_i-x_0)\right)^2}{\Delta F_i^2 + b^2 \Delta t_i^2},\label{chierr}\end{equation}
where $\Delta F_i = -f_i^{-11/3}\Delta f$, $b = (96/5)\pi^{8/3}\left(\text{G}M_c/\text{c}^3\right)^{5/3}$ and $x_0 = C/b$.
The value of $\chi^2$ measures the quality of the fitted line, which has the slope b and the intercept C. First the line of the best fit is estimated by maximizing the number of data points, which are intersected  
with the line. If a data point has a $\chi^2$ value of less than 2, it is counted towards the intersecting points. The value 2 has been hand picked. A larger value will increase the number of intersecting 
data points as it will also consider distant data points from Gaussian noise and hence may not achieve a good fit. A smaller value may try to fit a line through a region densely packed with data points and may fit line through  just the merger stage of the waveform.

Ones the fitting pixel set is defined, the line of the best fit is obtained by minimizing the $\chi^2$ value and the chirp mass is calculated from the slope b. The $\chi^2$ in Equation~\ref{chierr} is not 
normalized, as $\Delta F$ and $\Delta t$ are not statistical errors but the uncertainities in time and frequency. Hence, it cannot be used to estimate error in the reconstructed chirp mass. Alternatively, the 
goodness of fit is assessed by analyzing if an event has a chirp like structure. This is quantified using two parameters, 1) energy fraction (F) =  the energy of all pixels intersected by the line divided by 
the total energy of the event. 2) ellipticity (e) = 1 - $\chi^2/\chi^2_\bot$, where $\chi^2_\bot$ corresponds to the line orthogonal to the line of best fit. The maximum possible value of both the parameters is 
1 and a chirping GW signal is expected to have F and e close to unity. To impose a chirping signal criteria, only events crossing certain thresholds on reconstructed chirp mass, F and e are admitted for further 
investigation.

\section{Application of the chirp cut \label{app}}

Chirp cut was tested on the data collected by the LIGO and the Virgo detectors from the S6A-VSR2 run. The excess power events were identified with the coherent waveburst algorithm and background 
was estimated by performing the time-shift analysis i.e. events were identified after data from two detectors were shifted to remove the possibility of coincident GW signal \cite{kymm2008}. The 
sensitivity of the analysis to simulated CBC waveforms was estimated with Monte Carlo simulations by using waveforms from the EOBNRv2 family with a uniform distribution of chirp mass from 4.35 M$_\odot$ 
to 21.76 M$_\odot$ \cite{lvc2012}. Only events identified with a signal-to-noise ratio (SNR) of 6 or more were considered for further processing.

\begin{figure}[htbp]
 \begin{center} 
 \includegraphics[width=0.45\textwidth]{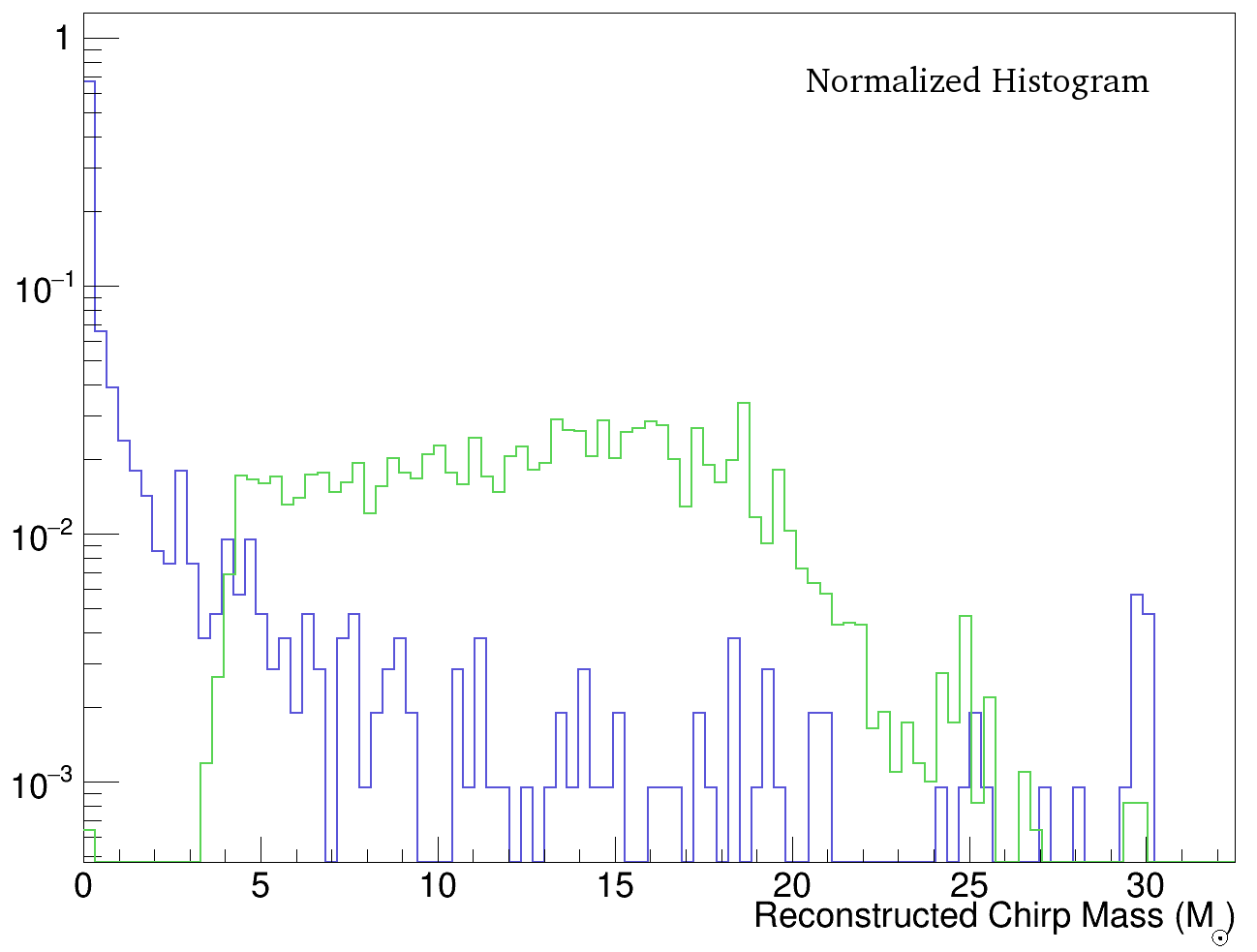}
 \end{center}
\caption{\label{rec_mchirp} Green histogram corresponds to reconstructed chirp mass values for the EOBNRv2 injections. The injections cover a chirp mass range from 4.35 M$_\odot$ to 21.76 M$_\odot$. Blue 
histogram corresponds to reconstructed chirp mass of background events for the same data. Most of the background events have been reconstructed with a chirp mass value of less than one solar mass. To make 
comparison easier, histograms have been normalized such that total number of events in a histogram is one.}
\end{figure}

The background events are unlikely to have a chirp structure. In fact for an event with data points randomly distributed on the TF map, the reconstructed chirp mass will be close to zero. Hence, 
the first cut is applied to the reconstructed chirp mass. Figure \ref{rec_mchirp} plots the distribution of reconstructed chirp mass for simulated signals (injections) and background events. A large 
fraction of the background events have a reconstructed chirp mass value of less than one solar mass. Setting a threshold of $M_c = 1$ on the reconstructed chirp mass value removes a large fraction of 
background events while the number of injection removed from this cut is negligible. 

\begin{figure}[htbp]
 \begin{center} 
 \includegraphics[width=0.45\textwidth]{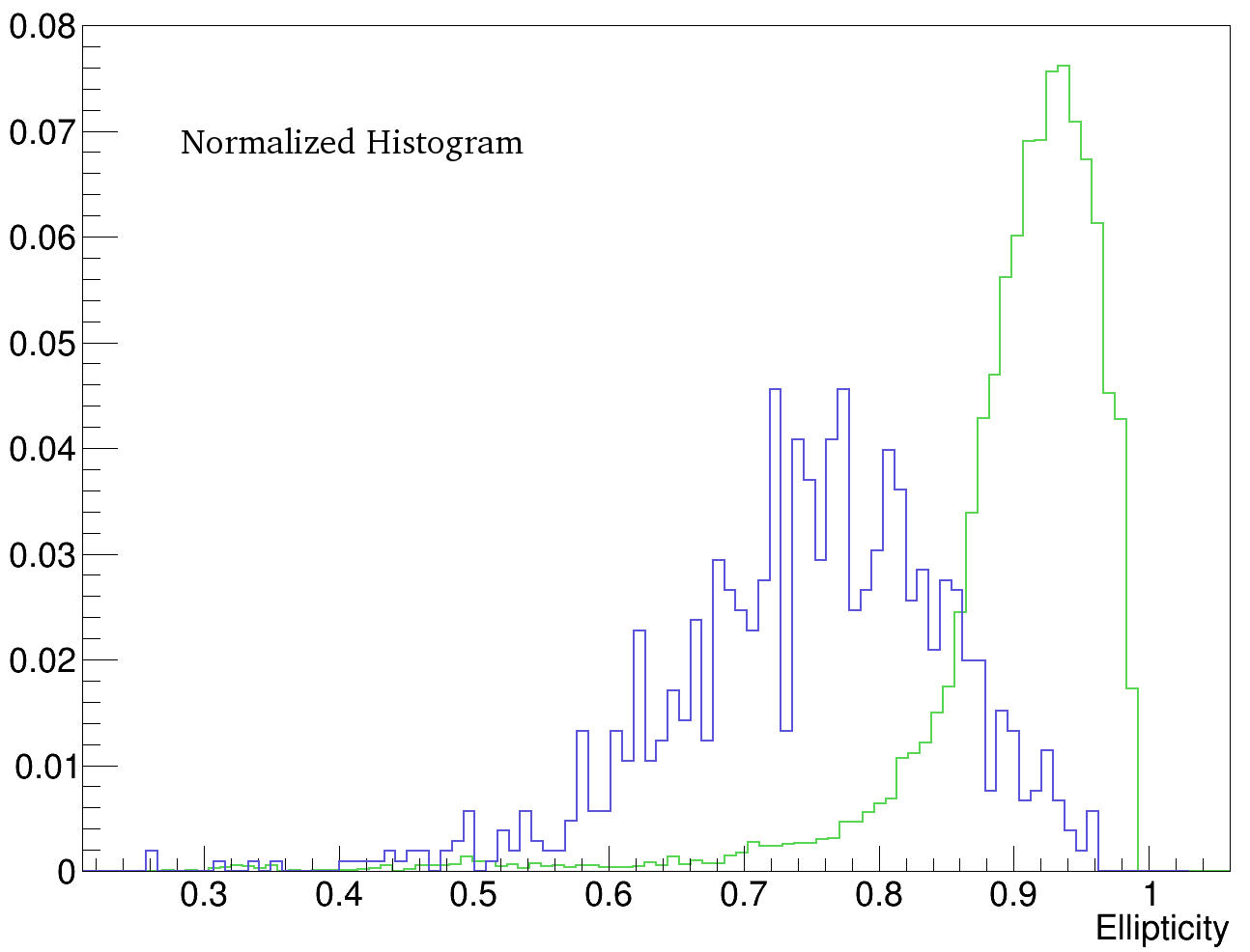}
 \end{center}
\caption{\label{ellip} Green histogram corresponds to the ellipticity distribution for injections and blue histogram corresponds to the ellipticity distribution for background events.}
\end{figure}

Events comprised of small number of data points may have a large value of reconstructed chirp mass. Such events can be vetoed by using their ellipticity. A chirping signal is expected to have e close to 
unity, while background events are expected to have value close to 0. Figure \ref{ellip} plots the distribution of ellipticity for the injections and background events. 
The ellipticity threshold of $e = .8$ has been used to produce the results presented in this letter.

\begin{figure}[htbp]
 \begin{center} 
 \includegraphics[width=0.45\textwidth]{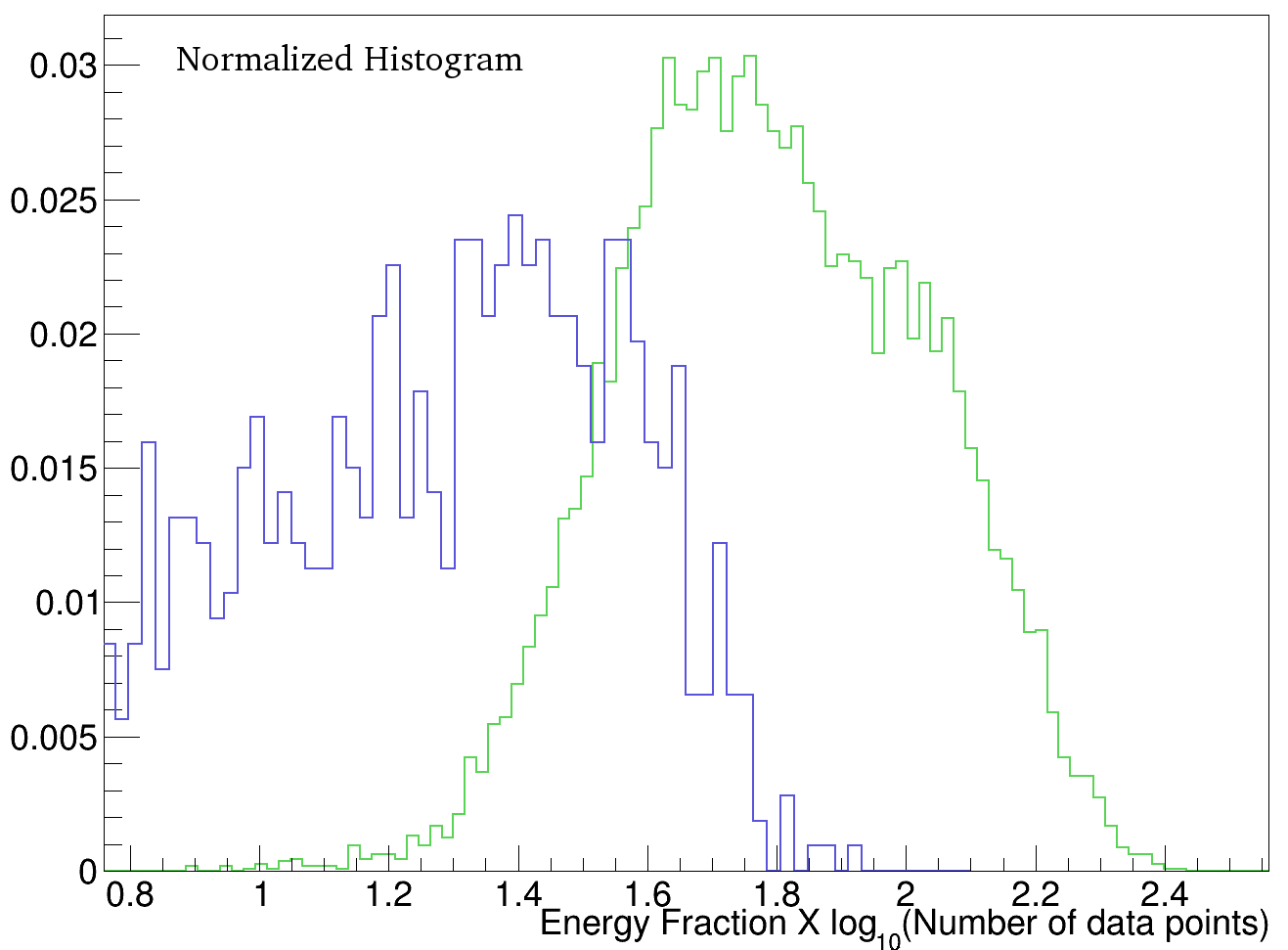}
 \end{center}
\caption{\label{en_frac} Green histogram corresponds to the energy fraction distribution (weighted by log of number of data points in an event) of EOBNRv2 injections made on the LHV S6A data. The injections 
cover a chirp mass range from 4.35 $M_\odot$ to 21.76 $M_\odot$. Blue histogram corresponds to the reconstructed chirp mass of background events.}
\end{figure}

Background events may have a large value of reconstructed chirp mass as well as ellipticity when number of data points in the event are large. Such events can be removed by using their energy fraction value. 
Events with small number of data points are not penalized by applying the cut on the energy fraction statistics weighted by $\log_{10}(\text{number of data points in the event})$. Figure \ref{en_frac} 
plots the distribution of weighted energy fraction for injections and background events. The energy fraction threshold $F = 1.3$ has been used in the results presented in this letter.

\section{Results \label{res}}

\begin{figure}[htbp]
 \begin{center} 
 \includegraphics[width=0.5\textwidth]{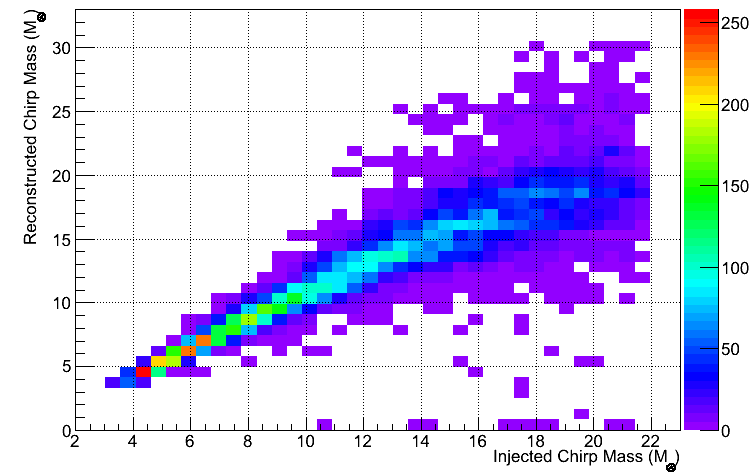}
 \end{center}
\caption{\label{inj_rec_mchirp} Plot shows the distribution of reconstructed vs injected chirp mass for the recovered injections. Recovered chirp mass is within few percent of the injected chirp mass. }
\end{figure}

Figure \ref{inj_rec_mchirp} plots the injected vs reconstructed chirp mass of all the recovered events. Reconstructed chirp mass is within a few percent of the injected chirp mass. The curve begins to 
flatten (20 M$_\odot$ and onwards) for higher values of injected chirp mass primarily because the inspiral stage becomes covered by the seismic noise affecting the detector. 

\begin{figure}[htbp]
 \begin{center} 
 \includegraphics[width=0.45\textwidth]{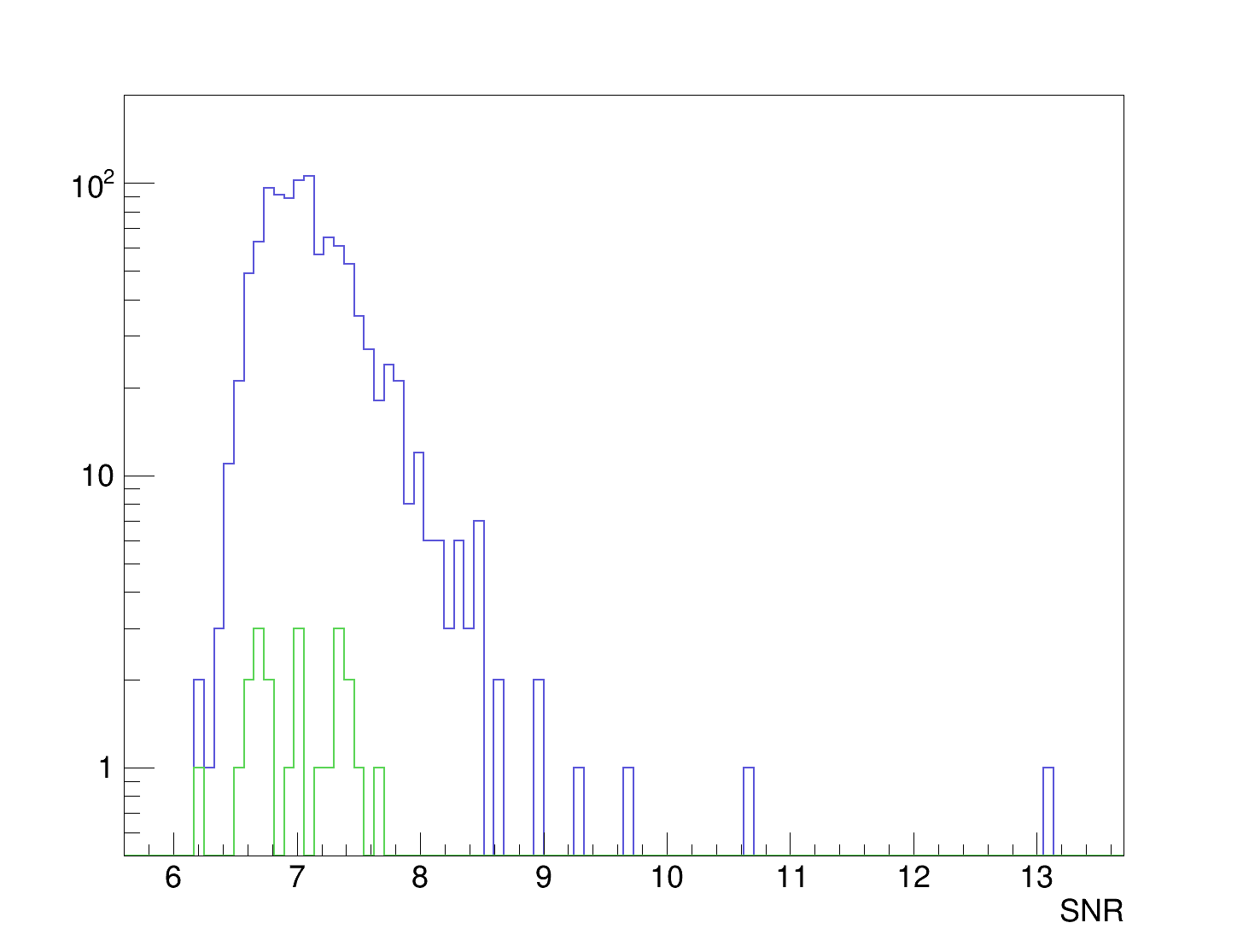}
 \end{center}
\caption{\label{bkg_red} Plot shows the distribution of background events before and after the application of chirp cut. Blue histogram shows the SNR distribution of the events obtained from the time-shift 
analysis and the event that survive the chirp cut are shown by the green histogram.}
\end{figure}

\begin{figure}[htbp]
 \begin{center} 
 \includegraphics[width=0.45\textwidth]{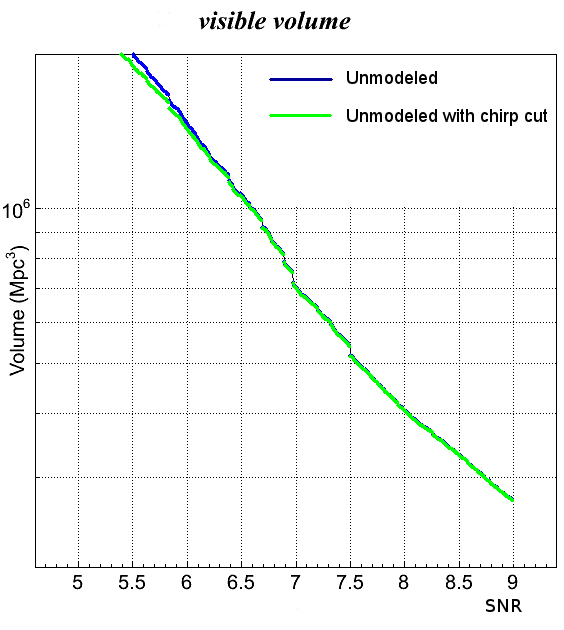}
 \end{center}
\caption{\label{vol_comp} Visible volume as a function of SNR. The visible volumes are calculated in terms of EOBNRv2 waveforms and averaged over the tested parameter space. Blue curve is obtained by 
considering all the recovered injections and green curve is obtained from the recovered injections surviving the chirp cut. The loss is sensitivity is negligible.}
\end{figure}

The primary objective of the chirp cut is to reduce the background while not affecting the search sensitivity. This is shown in Histogram \ref{bkg_red}, which plots the SNR distribution of the background 
events before and after the application of the chirp cut. There is a significant reduction in the number of background events (approximately two orders of magnitude) and the outliers have also been removed. 
Moreover, the sensitivity of the search remained mostly unchanged as shown in Figure \ref{vol_comp}. It plots the visible volume of the search before and after the application of chirp cut. The 
loss in visible volume is negligible, indicating that most of the recovered injections passed the chirp cut.

\section{Conclusion \label{conclusion}}

The chirp cut is a powerful tool to reduce background without using the CBC template analysis, which requires knowledge of the exact source model. The algorithm admits only events with chirping TF signature 
resulting in background reduction by two orders of magnitude. Other than the overall reduction of the background, algorithm also succeeds in removing most of the loud outliers. On the other hand it does not 
compromise the search sensitivity for the CBC sources, therefore its application will boost the statistical confidence of these sources. So far stellar mass CBC sources have been searched for with the matched 
filtering methods using simulated waveforms of binaries on circular orbits. However, it has been shown that these methods could be sub-optimal for sources such as eccentric binaries, intermediate mass-ratio 
coalescence or extreme cases of precessing binaries\cite{bz2010, sm2013, ia2014}, and the burst
search can be a viable alternative. 
%A matched filtering search can be setup for these sources but it will be computationally and scientifically challenging. 
%Burst search, being mostly insensitive to phase of the waveform, combined with chirp cut 
%is a  choice for these sources. The chirp cut has been employed in 
%the current eccentric binary search. Search has 
%been conducted on the data obtained from the LIGO/Virgo detectors using the coherent waveburst algorithm.
%\cite{t2015}. 

The strength of chirp cut lies in its simple application. Algorithm is not computationally intensive and does not require separate computational infrastructure when implemented in a search. Reconstructed 
chirp mass, ellipticity and energy fraction can be estimated during an unmodeled all-sky burst search and the chirp cut can be imposed once all the excess energy events have been identified. CBC signals 
are expected to last multiple seconds for the sensitivity of advanced detectors. We expect chirp cut to be an effective tool in search of GWs from the CBC sources. 

\section{Acknowledgement }

We are thankful to the  National Science Foundation for support 
under grants PHY 1205512 and PHY 1505308. This document has been assigned LIGO Laboratory document number P1500171.

%\clearpage

%\newpage
%\input{review1}

\end{document}